\documentclass[hyper]{JHEP} 

\usepackage{epsfig}

\newcommand\fverb{\setbox\pippobox=\hbox\bgroup\verb}
\newcommand\fverbdo{\egroup\medskip\noindent%
			\fbox{\unhbox\pippobox}\ }
\newcommand\fverbit{\egroup\item[\fbox{\unhbox\pippobox}]}
\newbox\pippobox
\title{The Schr\"odinger Wave Functional 
and S-branes}
\author{ J. Kluso\v{n}
\footnote{On leave from Masaryk University, Brno}\\
Institute of Theoretical Physics, University of Stockholm, SCFAB\\
SE- 106 91 Stockholm, Sweden \\
and \\
Institutionen f\"or teoretisk fysik\\
BOX 803, SE- 751 08 
Uppsala, Sweden \\
E-mail: \email{josef.kluson@teorfys.uu.se}}

\preprint{\hepth{0307079}}

\abstract{In this paper we 
 will consider the 
minisuperspace approach to S-branes
dynamics in the Schr\"odinger picture
description. Time-evolution of vacuum wave functional
for quantum field theory on S-brane 
is studied. Open string pair production is calculated. The
analysis of  density matrix for mixed
states is also performed.}

\def\tr{\mathrm{Tr}}

\def\bra #1{\left<#1\right|}
\def\ket #1{\left|#1\right>}

\def\bk{\mathbf{k}}
\def\bx{\mathbf{x}}
\def\by{\mathbf{y}}
\def\bz{\mathbf{z}}

\def\bp{\mathbf{p}}

\def\ss{\sin \frac{\tau}{\sqrt{2}}}
\def\st{\sinh \frac{\tau}{\sqrt{2}}}
\def\st2{\sinh^2 \frac{\tau}{\sqrt{2}}}
\def\ss2{\sin^2 \frac{\tau}{\sqrt{2}}}

\def\ik{\frac{d\bk}{(2\pi)^{p+1}}}
\def\tg{\tilde{G}}
\def\ma{\mathcal{A}}
\def\ikk{\frac{d\bk'}{(2\pi)^{p+1}}}
\def\ikkk{\frac{d\bk''}{(2\pi)^{p+1}}}

\def\hp{\hat{\pi}}
\def\hf{\hat{\phi}}
\begin{document}
\section{Introduction}\label{first}
In \cite{Gutperle:2002ai}
new string theory ingredients  called
S (spacelike)-branes were introduced. Whereas
ordinary D-branes  can be realised
as  timelike kinks and vortices of
the tachyon field 
\cite{Sen:1999mg,Witten:1998cd,Horava:1998jy},
 spacelike defects can be defined as spacelike
kinks and vortices in the background of
a time-dependent tachyon condensation process
called rolling tachyon
\cite{Gutperle:2002ai,Sen:2002an,
Sen:2002in,Sen:2002nu}.
 These 
S-branes can be thought of as the creation
and subsequent decay of an unstable brane. This process
has recently attracted much attention
\cite{Gutperle:2002ai,Sen:2002an,
Sen:2002in,Sen:2002nu,Strominger:2002pc,Gutperle:2003xf,
Maloney:2003ck,Hashimoto:2003qx,Hashimoto:2002sk,
Sen:2002qa,
Kutasov:2003er,Lambert:2003zr,McGreevy:2003kb,
Moeller:2003gg,Gaiotto:2003rm,
Aref'eva:2003qu,Yang:2002nm,Okuda:2002yd,
Mukhopadhyay:2002en,Moeller:2002vx,Sen:2002vv,
Minahan:2002if,Sugimoto:2002fp,Kluson:2003xu,Kluson:2002av,
Fujita:2003ex,Berkooz:2002je,Felder:2002sv,Frolov:2002rr,
Burgess:2002vu,Rey:2003xs,Rey:2003zj,
Klebanov:2003km,McGreevy:2003ep,Constable:2003rc,
Demasure:2003av,Martinec:2003ka,Gubser:2003vk,
Kwon:2003qn,Karczmarek:2003,
Sen:2003od,Suqawara:2003tp,Ohmori:2003,Kluson:2003sb}. 
S-branes have been also extensively
studied in supergravity approach with potentially 
interesting cosmological applications
\cite{Kruczenski:2002ap,Chen:2002yq,
Leblond:2003db,Burgess:2003gg,Ohta:2003uw,
Gutperle:2003kc,Ohta:2003ie,Roy:2003nd,Ohta:2003pu,
Piao:2002nh,Shiu:2002cb,Buchel:2002tj,Buchel:2002kj,
Li:2002et,Mukohyama:2002cn,
Buchel:2003xa,Leblond:2003ac,McInnes:2003vu,
Roy:2003ra,Piao:2002vf,Guo:2003zf,
Deger:2002ie,Deger:2003fz}.

Very nice worldsheet construction of S-branes was given
in the classical $g_s=0$ limit by A. Sen
\cite{Sen:2002an,Sen:2002in,Sen:2002nu} where
he introduced class of models in bosonic string theory
obtained by perturbing the flat space $c=26$ 
$CFT$ with the exactly marginal deformation
\begin{equation}\label{bound1}
S_{boud}=\lambda \int d\tau \cosh X^0(\tau) \ ,
\end{equation}
where $X^0$ is time coordinate, $t$ is a coordinate on the
worldsheet boundary and $\lambda$ is a free parameter
in the range $0\geq \lambda \leq \frac{1}{2}$
\footnote{We work in units $\alpha'=1$.}. This is family
of exact solutions of classical open string theory
whose space-time interpretation is that of an unstable brane
being created at a time $X^0 \sim -\tau$ and decaying
at a time $X^0 \sim \tau$ with $\tau=-\log (\sin (\pi\lambda))$. 

As was shown in  
\cite{Strominger:2002pc,Gutperle:2003xf,Maloney:2003ck} 
this time-dependent process of tachyon condensation that
in fact defines S-brane has many intriguing properties. 
In particular, it is known that in time-dependent backgrounds
there is in general no preferred vacuum and particle production
is unavoidable. In \cite{Maloney:2003ck}
the open string vacua on S-brane were studied. It was shown
that for (\ref{bound1}) there is open string pair
production with a strength characterised by the Hagedorn
temperature $T_H=\frac{1}{4\pi}$ \cite{Strominger:2002pc,
Gutperle:2003xf}. As was argued 
\cite{Strominger:2002pc,Gutperle:2003xf,Maloney:2003ck} 
this temperature arises from the periodicity of the boundary
interaction (\ref{bound1}) in imaginary time
\footnote{The appearance of Hagedorn temperature signals
a breakdown of string perturbation theory. 
To avoid this problems we can work at the
limit of vanishing string coupling constant $g_s=0$.}. 

In \cite{Strominger:2002pc,Gutperle:2003xf,Maloney:2003ck} 
S-brane vacua and their properties were studied in
minisuperspace approximation where all open string modes
acquire exponentially growing masses from tachyon background.
The result is open string pair production. It was shown in
\cite{Strominger:2002pc} that for any nonzero $g_s$ 
this production is typically divergent.  The divergent density
of open strings with exponentially growing masses will couple
strongly to closed strings suggesting that S-brane density
is quickly released into closed strings. 

The  minisuperspace description of S-branes  studied in
\cite{Strominger:2002pc,Gutperle:2003xf,Maloney:2003ck} 
was performed in the
Heisenberg picture description of quantum field theory  where we treat quantum
fields as Heisenberg operators that are time-dependent
 while quantum states do not evolve. 
As companion to this approach  we will analyse the minisuperspace approach
of  S-brane using the Schr\"odinger formulation 
 of quantum field theory.  The Schr\"odinger picture 
 of quantum field theory provides a simple and
intuitive description of vacuum states in quantum field
theory in situations where the background metric
is time-dependent or in case where some parameters
of quantum field theory depend on time. The Schr\"odinger
picture characterises vacuum states explicitly by  simple
wave functional specified single, possibly time-dependent,
kernel function satisfying a differential equation with
prescribed boundary conditions. This makes no reference
to the assumed spectrum of excited states and so circumvents
the difficulties of the conventional canonical description
of a vacuum as a "no-particle" state with respect to the
creation and annihilation operators defined by a particular
mode decomposition of the field, an approach which is not
well suited to the time-dependent problems as for example
minisuperspace description of S-branes. 

The wave functional that defines the vacuum state satisfies
a functional Schr\"odinger equation describing its time evolution.
 We will explicitly  construct 
Gaussian vacuum state functional for quantum field
theory on half S-brane and on S-brane.
We choose the initial condition in such a way
that the vacuum wave  functional  approaches
standard Gaussian flat space functional in 
the far past $t\rightarrow -\infty$ in case of half S-brane,
or that approaches Minkowski flat space vacuum
functional for $t\rightarrow 0$ in case of S-brane.  Then we
will study the aspect of particle production
during time evolutions of these vacuum states.
Next we turn to the S-brane thermodynamics. We will construct
the density matrix for mixed states. Since the
rolling tachyon background is time-dependent process 
the S-branes are highly time-dependent configurations.
However temperature is an equilibrium (or at best 
adiabatic) concept, so it is usually does not
to make sense to put a time-dependent configuration
at finite temperature.  The standard procedure applied
to the half S-brane  is to
assume that at time $t_0=-\infty$ the initial
density matrix is thermal with the temperature
$T=1/\beta$. It is only in this initial state
that the notion of temperature is meaningful. As the system
departs from equilibrium one cannot define
a thermodynamic temperature. We will construct these 
mixed states that approach standard thermal 
vacuum at past infinity. Following 
 \cite{Maloney:2003ck} we will show that
there is sort of mixed states with temperature
$T=\frac{1}{2\pi n}$ that retain thermal 
periodicity at all times. For full S-brane we will
proceed in the same way however we will
demand that density matrix approaches standard
thermal density matrix at  $t_0=0$ when 
 in the limit of  $\lambda \rightarrow 0$ 
the interaction can be neglected and we have
ordinary unstable D-brane. 

This paper is organised as follows. In the next
section (\ref{second})
we   review minisuperspace analysis
of S-branes, following  \cite{Maloney:2003ck}.
We will demonstrate on an example of half S-brane
the  fact that particle creation is natural
process during unstable D-brane decay.
In section (\ref{third}) we will study 
the Schr\"odinger picture description of
quantum field theory on half S-brane. 
 We will explicitly
construct the vacuum wave functional 
that approaches standard Minkowski vacuum wave
functional at the asymptotic past $t\rightarrow -\infty$.
We will calculate the number of particle produced during
time evolution of this state and confirm the result
\cite{Maloney:2003ck} that open string particle production
is natural process during D-brane decay. In
section (\ref{fourth}) we will apply Schr\"odinger
picture description to the full S-brane. In section (\ref{fifth}) 
we will study S-brane thermodynamics. 
We will 
construct density matrix for mixed states
in Schr\"odinger approach for half S-brane and
for full S-brane. 
 We will also  calculate the equal-time correlators in
these mixed states. 
And finally in  conclusion (\ref{seventh}) we outline our results given
in this paper.

\section{Review of the minisuperspace approach}\label{second}
In this section we present a short review of superspace
approach to the study of S-brane dynamics 
\cite{Strominger:2002pc,Gutperle:2003xf,Maloney:2003ck}. We will
closely follow these papers.

We wish to understand the dynamics of the open string worldsheet
theory with a time-dependent tachyon
\begin{equation}\label{action}
 S=-{1\over
4\pi}\int_{\Sigma_2} d^2\sigma \partial^a X^\mu \partial_a X_\mu + \int_{
\partial 
\Sigma_2} d\tau \ m^2(X^0) \ .
\end{equation}
 For the open bosonic string $m^2=T$
where $T$ is the spacetime tachyon, while for the open superstring
$m^2 \sim T^2$ after integrating out worldsheet fermions. We use
the symbol $m^2$ to denote the interaction because the coupling
(among other effects) imparts a mass to the open string states. We
consider three interesting cases described by the marginal
interactions
\begin{eqnarray}
 m^2_+(X^0) = \frac{\lambda}{2} e^{X^0}  
\\
m^2_-(X^0) = {\lambda \over 2} e^{-X^0} 
\\
 m^2_s(X^0)
= \l \cosh X^0  
\end{eqnarray}
The first case $m_+^2$ describes the
process of brane decay, in which an unstable brane decays via
tachyon condensation. The second case describes the time-reverse
process of brane creation, in which an unstable brane emerges from
the vacuum. The final case describes an S-brane, which is
the process of brane creation followed by brane decay. Brane decay
 can be thought of as the future (past) half of an
S-brane, i.e. as the limiting
case where the middle of the S-brane is pushed into the infinite past
(future).
 
We will  study S-brane dynamics using the minisuperspace
analysis  in which the effect of the interaction is simply to
give a time-dependent shift  to the masses of
all the open string states.   
In the minisuperspace approximation
only the zero-mode dependence of the interaction $m^2(X^0)$ is
considered.  In this case we can plug in the usual mode
solution for the free open string with oscillator number $N$ to
get an effective action for the zero modes 
\begin{equation}
 S = \int
d\tau \left(-\frac{1}{4}\dot{x}^\mu \dot{x}_\mu + (N-1) +
2m^2(x^0)\right)  \ .
\end{equation}
 This is the action of a point particle with a
time dependent mass. Here $x^\mu (\tau)$ is the zero mode part of
$X^{\mu} (\sigma,\tau)$, and the second term in  is an effective
contribution from the oscillators, including the usual normal
ordering constant. From upper action we can write down the Klein-Gordon
equation for the open string wave function $\phi(t,\bx)$,
\begin{equation}
 \left(\partial^\mu\partial_\mu - 2{m^2(t)} -({N-1})
\right)\phi(t,\bx)=0, 
\end{equation}
 where $(t, \bx)$ are the spacetime
coordinates corresponding to the worldsheet fields $(X^0,X^i)$.
 This is the equation of motion for a scalar field with
time-dependent mass.
At this point, we should make a few remarks about field theories
with time-dependent mass
\footnote{For review of quantum field theory in curved
spacetime, see \cite{Birrell:ix,Fulling:nb,Wald:yp}.}.
 Time translation invariance has been
broken, so energy is not conserved and there is no preferred set
of positive frequency modes. This is  a familiar circumstance in
the study of quantum field theories in time-dependent backgrounds
which leads to particle creation. The probability current
$j_\mu= i (\phi^*\partial_\mu \phi - \partial_\mu \phi^* \phi)$ is 
still
conserved, allowing us to define the Klein-Gordon inner product
\begin{equation}\label{kgcur}
\langle f|g\rangle = i \int_\Sigma d \Sigma^\mu
(f^*\partial_\mu g - \partial_\mu f^* g) 
\end{equation}
 where $\Sigma$ is a spacelike
slice.  This norm does not depend on the choice of $\Sigma$ if $f$
and $g$ solve the wave equation. Normalised positive frequency
modes are chosen to have $\langle f|f\rangle=1$. Negative
frequency modes are complex conjugates of positive frequency
modes, with $\langle f^*| f^*\rangle=-1$. There is a set raising
and lowering operators associated to each choice of mode
decomposition -- these operators obey the usual oscillator algebra
if the corresponding modes are normalised with respect to (\ref{kgcur}).
We also define a vacuum state associated to each mode
decomposition -- it is the state annihilated by the corresponding
lowering operators.
To illustrate this idea we give an example of
the half S-brane corresponding to the decay of 
an unstable D-brane. In other words, we will consider 
previous equation with the mass term
\begin{equation}
m^2_+(t)=\frac{\lambda}{2}e^t \ .
\end{equation}
Since in the far past $t\rightarrow -\infty$ the mass term $m^2_+(-\infty)=0$,
it is natural to take all open string modes in their usual
ground state. In other words, we define 
$\ket{in}$ vacuum as a vacuum with no particle present.
Expanding field in plane waves 
\begin{equation}
\phi_{\bp}(x)\equiv\phi_{\bp}(t,\bx)=e^{i\bp\bx}
u_{\bp}(t) 
\end{equation}
the wave equation becomes
\begin{equation}
(\partial_t^2+\lambda e^t+\omega^2_{\bp})u_{\bp}=0 \ ,
\omega^2_{\bp}=p^2+N-1\equiv p^2+m^2 \ . 
\end{equation}
This is a form of Bessel's equation. It has normalised,
positive frequency solutions 
\begin{equation}\label{uin}
u^{in}_{\bp}=\lambda^{i\omega_{\bp}}\frac{
\Gamma(1-2i\omega_{\bp})}
{\sqrt{2\omega_{\bp}}}J_{-2i\omega_{\bp}}
(2\sqrt{\lambda}
e^{t/2}) \ , 
\end{equation}
where superscript $in, out$ and $0$ on a wave function 
denotes solutions that are purely positive frequency when
$t\rightarrow -\infty,t\rightarrow 0$ or $t=0$. 
These solutions have been chosen because they approach flat
positive frequency plane waves in the far past $t\rightarrow
-\infty$ 
\begin{equation}
u^{in}_{\bp}\sim \frac{1}{\sqrt{2\omega_{\bp}}}
e^{-i\omega_{\bp} t} \  . 
\end{equation}
We will also consider the wave functions
\begin{equation}\label{uout}
u^{out}_{\bp}=\sqrt{\frac{\pi}{2}}
(ie^{2\pi\omega_{\bp}})^{-1/2}H^{(2)}_{-2i\omega_{\bp}}
(2\sqrt{\lambda}e^{t/2}) \ ,
\end{equation}
that are purely positive frequency in the far future
$t\rightarrow \infty$
\begin{equation}
u^{out}_{\bp}\sim \frac{\lambda^{-1/4}}{
\sqrt{2}}\exp \left\{-t/4-2i\sqrt{\lambda}e^{t/2}\right\} \ . 
\end{equation}
Using these solutions we can define  vacuum state 
in general time-dependent background. In the canonical
framework we perform an
expansion of scalar field in terms of modes that are solutions of
free wave equation and that have positive and negative norm
with respect Klein-Gordon product. Then  operators that come
with positive modes are regarded as lowering operators that
annihilate vacuum state and operators that come with negative
norm modes are rising operators. More precisely, we have 
an expansion
\begin{equation}
\phi(x)=\sum_{\bp} a_{in,\bp}\phi^{in}_{\bp}(x)+
a_{in,\bp}^{\dag}\phi^{in*}_{\bp}(x) \ , 
\end{equation}
where $\phi^{in}(x)=e^{i\bp\bx}u^{in}_{\bp}(t)$ and
$u^{in}_{\bp}(t)$ are given  in (\ref{uin}). 
Then   $\ket{in}$ vacuum  state is defined as  the state that is annihilated by
all $a_{in,\bp}$
\begin{equation}
a_{in,\bp}\ket{in}=0 \ .
\end{equation}
In the same way we can write
\begin{equation}
\phi(x)=\sum_{\bp} a_{out,\bp}\phi^{out}_{\bp}(x)+
a_{out,\bp}^{\dag}\phi^{out *}_{\bp}(x) \ ,
\end{equation}
where $\phi^{out}_{\bp}(x)=
e^{i\bp\bx}u^{out}_{\bp}(t)$ and
$u^{out}_{\bp}(t)$
 are given in
(\ref{uout}). Now we 
define $\ket{out}$ vacuum state  as the state that
is annihilated by all $ a_{out, \bp}$
\begin{equation}
a_{out,\bp}\ket{out}=0 \ .
\end{equation}
Generally $u^{out}_{\bp}$ and $u^{in}_{\bp}$ are related by 
celebrated Bogolubov transformations
\begin{equation}\label{gom}
u^{out}_{\bp}=A_{\bp}
u^{in}_{\bp}+B_{\bp}u^{in*}_{\bp} \ ,  
\end{equation}
where  
\begin{equation}\label{bol}
A_{\bp}=e^{2\pi \omega_{\bp}
+\pi i/2}B^*_{\bp}=
\sqrt{\omega_{\bp}}e^{\pi\omega_{\bp}
-\pi i/4}\left(
\frac{\lambda^{-i\omega_{\bp}}}{\sinh 2\pi 
\omega_{\bp} 
\Gamma(1-2i\omega_{\bp})}\right) \ . 
\end{equation}
These coefficients obey unitarity relation
$|A_{\bp}|^2-|B_{\bp}|^2=1$. 
From  (\ref{gom}) we get  the relation between
in and out creation and annihilation operators
\begin{equation}
a_{in,\bp}=A_{\bp}a_{out, \bp}+B^*_{\bp}
a_{out,\bp}^{\dag} \ . 
\end{equation}
From this, the condition $a_{in, \bp}\ket{in}=0$ implies
that $\ket{in}$ is squeezed state 
\begin{equation}
\ket{in}=\prod_{\bp}(1-|\gamma_{\bp}|^2)^{1/4}
\exp\left\{-\frac{1}{2}\gamma_{\bp}
(a_{out,\bp}^{\dag})2\right\}\ket{out} \ ,
\gamma_{\bp}=B_{\bp}^*/A_{\bp} \ .
\end{equation}
Physically this 
means that particles are produced during
brane decay: if we start in a state with no particles
at $t\rightarrow -\infty$, there will be many particles
at time $t\rightarrow \infty$ with the density of 
particles with momentum $\bp$ 
\begin{equation}\label{nk}
\mathcal{N}_{\bp}=|B_{\bp}|^2=\frac{1}{e^{\omega_{\bp}/T_H}-1} \ ,
T_H=\frac{1}{4\pi} \ .
\end{equation}
Despite the fact that $\ket{in}$ is pure state the particle
density at far future is the same as thermal density of particles 
 at temperature $T_H=1/4\pi$. In string units, 
$T_H$ is the Hagedorn temperature. As was argued 
in \cite{Strominger:2002pc,
Maloney:2003ck,Lambert:2003zr,Okuda:2002yd} the 
appearance of the Hagedorn temperature signals a breakdown
of string perturbation theory. To find mechanisms which
cuts off this divergence is interesting problem which
we will not try to solve in this paper. To avoid this
problem we will work, following \cite{Maloney:2003ck}
at $g_s=0$.

In the next section we will formulate the Schr\"odinger
picture description of the quantum field theory on
S-brane which is especially useful for the study
of quantum field theory in time-dependent background. 
\section{QFT in Schr\"odinger picture}\label{third}
In this section we  introduce 
Schr\"odinger formalism for QFT on half
S-brane. We  follow mainly 
 \cite{Long:1996wf,Guven:1987bx,Eboli:1988qi,
Eboli:1988pr}.
We begin with   an  action for
 massive scalar field  which
describes open string modes on Sp-brane
 in the minisuperspace approach
\footnote{Our convention is
$\eta_{\mu\nu}=\mathrm{diag}(-1,\dots, 1) \ ,
\mu\ , \nu=0,\dots, p+1$, $a,b,c,\dots=1,\dots,p+1$.
We also denote $x^0=t$ and $\bx=(x^1,\dots,x^{p+1})$.}
\begin{equation}
S=\int dtd\bx L=
-\frac{1}{2}\int dtd\bx\left(
-\partial_t\phi\partial_t\phi+\eta^{ab}
\partial_a\phi\partial_b\phi+m^2(t)\phi^2
\right) \ .
\end{equation}
The canonical momentum conjugate to $\phi(t,\bx)$ is
\begin{equation}
\pi(t,\bx)=\frac{\delta L}{\delta \partial_t\phi(\bx,t)}
=\partial_t\phi(t,\bx) \ 
\end{equation}
and the Hamiltonian
\begin{equation}
H=\int d\bx\left(\pi \partial_t\phi-L\right)=
\frac{1}{2}\int d\bx
\left(\pi^2+\eta^{ab}\partial_{a}\phi
\partial_b\phi+m^2\phi\right) \ .
\end{equation}
The system can be canonically quantized by treating the
fields as operators and imposing appropriate commutation
relations. This involves choice of a foliation
of a spacetime in a succession of
spacelike hypersurfaces. We choose these to
be the hypersurfaces of fixed $t$ and
impose equal-time commutation relations 
\begin{equation}
[\hf(\bx,t),\hp(\by,t)]=
i\delta (\bx-\by) \ , [\hf(\bx,t),\hf(\by,t)]=
[\hp(\bx,t),\hp(\by,t)]=0 \ .
\end{equation}
In the Schr\"odinger picture we take the basis vector of the state 
vector space to be the eigenstate of the field operator $\hf(t,\bx)$
on a fixed $t$ hypersurface, with eigenvalues $\phi(\bx)$
\begin{equation}
\hf(t,\bx)\ket{\phi(\bx),t}=
\phi(\bx)\ket{\phi(\bx),t} \ .
\end{equation}
Notice that the set of field eigenvalues $\phi(\bx)$
is independent of the value of $t$ labelling the
hypersurface.  In this picture, the quantum states
are explicit functions of time and are represented
by wave functionals $\Psi[\phi(\bx),t]$. Operators
$\hat{\mathcal{O}}(\hp,\hf)$ acting on these
states may be represented by
\begin{equation}
\hat{\mathcal{O}}(\hp(\bx),\hf(\bx))=
\mathcal{O}\left(-i\frac{\delta }{\delta
\phi(\bx)},  \phi(\bx)\right) \ .
\end{equation}
The Schr\"odinger equation which governs the
time evolution
of the wave functional  is
\begin{eqnarray}\label{schrod}
i\frac{\partial \Psi[\phi,t]}
{\partial t}=H\left(-i\frac{\delta }{\delta
\phi(\bx)},  \phi(\bx)\right)\Psi[\phi,t]=\nonumber \\
=\frac{1}{2}\int
d\bx \left[-\frac{\delta^2}{\delta \phi^2}
+\eta^{ab}\partial_a\phi\partial_b\phi+
m^2(t)\phi^2\right]\Psi[\phi,t] \ . 
\nonumber \\
\end{eqnarray}
To solve this equation, we make the ansatz, that up
to time-dependent phase, the vacuum functional is simple
Gaussian.  We therefore write
\begin{equation}\label{vacans}
\Psi_0[\phi,t]=
N_0(t)\exp\left\{-\frac{1}{2}\int
d\bx d\by \phi(\bx)G(\bx,\by,t)\phi(\by)\right\}=
N_0(t)\psi_0(\phi,t) \ .
\end{equation}
After inserting (\ref{vacans}) into
(\ref{schrod})  and comparing the terms of order $\phi^0$
we get 
\begin{equation}
i\frac{\partial N_0(t)}{\partial t}=
\frac{N_0(t)}{2}\int d\bz
G(\bz,\bz,t) \ .
\end{equation}
Comparing terms of order $\phi^2$ we obtain
\begin{eqnarray}\label{kernel}
i\frac{\partial G(\bx,\by,t)}
{\partial t}=
\int d\bz  G(\bz,\bx,t)G(\by,\bz,t)
-\left(\eta^{ab}\partial_a \partial_b
+m^2(t)\right)\delta(\bx,\by) \ .
\nonumber \\
\end{eqnarray}
Since  
the spatial sections on S-branes are flat it is natural
to perform a Fourier transformation on
the space dependence of the field configuration
\begin{equation}
\phi(\bx)=\int \ik e^{i\bk \bx}\alpha(\bk) \ .
\end{equation}
Similarly we can define $\delta/\delta \alpha(\bk)$
as 
\begin{equation}
\frac{\delta }{\delta \phi(\bx)}=
\int \ik e^{i\bk \bx}\frac{\delta }{
\delta \alpha(\bk)} \ .
\end{equation}
Reality of $\phi$ implies $\alpha^*(\bk)=
\alpha(-\bk)$. By definition the functional
derivative obeys 
\begin{equation}
\frac{\delta \phi(\bx)}{
\delta \phi(\bx')}=
\delta (\bx-\bx')
\end{equation}
which implies
\begin{equation}
\frac{\delta \alpha(\bk)}{
\delta \alpha(\bk')}=
(2\pi)^{p+1}\delta
(\bk+\bk') \ . 
\end{equation}
Functional integrals in the k-space 
variables require extra care because of
the constraint $\alpha^*(\bk)=\alpha(-\bk)$.
The path integral measure $\int d\phi$
is rewritten as
\begin{equation}
\int d\alpha(\bk)d\alpha^*(\bk)
\delta(\alpha(-\bk)-\alpha^*(\bk)) \ . 
\end{equation}
In the expression which results, it is 
simplest to perform all partial 
functional derivatives before carrying out
the functional integrations. Then
$\alpha^*(\bk)$ integration will
simply kill the $\delta$ function and
replace $\alpha^*(\bk)$ integration
by $\alpha(-\bk)$. 
In $k$-space, the Hamiltonian is
\begin{equation}\label{hamk}
H=\frac{1}{2}\int
\ik \left[-\frac{\delta^2}{\delta \alpha(\bk)
\alpha(-\bk)}+
\Omega^2_{\bk}(t)\alpha(\bk)\alpha(-\bk)\right] 
\ ,
\end{equation}
where
\begin{equation}
\Omega^2_{\bk}(t)=(m^2(t)-m^2)+\omega^2_{\bk} \ ,
\omega^2_{\bk}=k^2+m^2 \ .
\end{equation}
As expected in a free theory, a complete
decoupling of modes has been achieved
by the Fourier transformation.
In fact, for each $\bk$, the integrand
in (\ref{hamk}) represents
a harmonic oscillator with the time
dependent frequency $\Omega^2_{\bk}(t)$. 
After performing the Fourier transformation
for the kernel $G(\bx,\by,t)$
\begin{equation}
G(\bx,\by,t)=\int \ik
e^{i\bk(\bx-\by)}\tg(\bk,t) \ 
\end{equation}
the   kernel equation (\ref{kernel}) 
reduces to 
\begin{equation}
i\frac{\partial \tg(\bk,t)}
{\partial t}=\tg^2(\bk,t)-
\Omega^2_{\bk}(t) 
\end{equation}
that  can be solved with the
ansatz
\begin{equation}\label{kersol}
\tg(\bk,t)=-i\frac{\dot{\psi}_{\bk}(t)}{
\psi_{\bk}(t)} \ , \dot{f}=\frac{df(t)}{dt} \ , 
\end{equation}
where  $\psi_{\bk}(t)$ obeys
\begin{equation}\label{psieq}
\ddot{\psi}_{\bk}+\Omega^2_{\bk}(t)
\psi_{\bk}=0  \ .
\end{equation}
Now the vacuum state functional  has
the form
\begin{equation}
\Psi_0[\phi,t]=N_0(t)
\exp \left(\frac{i}{2}\int \ik 
\alpha(-\bk)\frac{\dot{\psi}_{\bk}(t)}
{\psi_{\bk}(t)} 
\alpha(\bk)\right) \ ,
\end{equation}
where 
\begin{equation}
\dot{N}_0(t)=
-i\frac{N_0(t)}{2}V\int \ik \tg(\bk,t)
\Rightarrow
N_0(t)=Ne^{-i\int^t dt'E_0(t')} \ 
\end{equation}
and
\begin{equation}
E_0(t)=\frac{1}{2}
V \int \ik \tg(\bk,t) \ .
\end{equation}
In the previous expression  
$V$ is  volume of spatial section on D(p+1)-brane and
 $N$ is time-independent normalisation constant. 
It is important to stress that (\ref{psieq})
is differential equation of the second order so that
its general solution is linear 
superposition  of two independent solutions and
hence it is parameterised by two independent constants.
On the other hand $\psi_{\bk}$ appears in $\tg_{\bk}$ 
in combination $\frac{\dot{\psi}_{\bk}}{\psi_{\bk}}$ 
 so that the overall normalisation factor 
is unimportant. As a result the vacuum  functional
contains one free parameter that 
is usually determined by   initial conditions we
impose on the vacuum wave functional.

Now we apply  the Schr\"odinger picture
description  reviewed
above to the case of the minisuperspace description
of  half S-brane. As we have seen in the
previous section this approach leads to the
equation of motion with time-dependent mass
and consequently 
the time-dependent frequency that
appears in 
(\ref{psieq}) is equal to
\begin{equation}
\Omega^2_{\bk}(t)=\omega_{\bk}^2+\lambda e^t \  .
\end{equation}
The general  solution of (\ref{psieq}) 
is 
\begin{equation}\label{psiin}
\psi_{\bk}(t)=A_{\bk}\psi^{in}_{\bk}(t)+
B_{\bk}\psi^{in *}_{\bk}(t) \ ,
\end{equation}
where \cite{Maloney:2003ck}
\begin{equation}
\psi_{\bk}^{in}(t)=
\lambda^{i\omega_{\bk}}\frac{
\Gamma(1-2i\omega_{\bk})}{\sqrt{
2\omega_{\bk}}}J_{-2i\omega_{\bk}} 
(2\sqrt{\lambda}e^{t/2}) \ . 
\end{equation}
To fix the  free parameter in the vacuum wave
functional we demand that in 
the  
asymptotic past  $t\rightarrow -\infty$ 
$\Psi_0[\phi,t]$ approaches the usual positive frequency 
Minkowski vacuum state 
\begin{equation}
\psi_0(\phi,-\infty)=
\exp\left(-\frac{1}{2}\int \ik
|\alpha(\bk)|^2\omega_{\bk}\right) \ .
\end{equation}
This initial condition   implies that
$A_{\bk}=0, B_{\bk}=1$.
We denote the vacuum state functional for QFT on
half  S-brane that approaches Minkowski vacuum  in
asymptotic past as $\Psi^{in}[\phi,t]$. 

Now we would like to examine the behaviour of
 $\Psi^{in}[\phi,t]$ 
 in the asymptotic future $t\rightarrow \infty$. It is convenient to
  express the kernel $\tilde{G}(\bk,t)=
-i\frac{\dot{\psi}^{in *}_{\bk}(t)}{\psi^{in *}_{\bk}(t)}$
in terms of the solution of 
(\ref{kernel})  $\psi^{out}_{\bk}$ 
\begin{equation}
\psi^{out}_{\bk}(t)=
\sqrt{\frac{\pi}{2}}(ie^{2\pi\omega_{\bk}})^{-1/2}
H^{(2)}_{-2i\omega_{\bk}}(2\sqrt{\lambda}
e^{t/2}) \ ,
\end{equation}
with simple  asymptotic behaviour 
for $t\rightarrow \infty$
\begin{equation}
\psi_{\bk}^{out}\sim
\frac{\lambda^{-1/4}}{\sqrt{2}}
\exp\left(-t/4-2i\sqrt{\lambda}
e^{t/2}\right) \ .
 \end{equation}
The relation between $\psi^{in}$ and $\psi^{out}$
modes is given by Bogolubov
transformation
\begin{eqnarray}
\psi^{out}_{\bk}=A_{\bk}\psi^{in}_{\bk}+
B_{\bk}\psi^{in*}_{\bk} \ , \nonumber \\
\psi^{in*}_{\bk}=A_{\bk}\psi^{out*}_{\bk}-
B^*_{\bk}\psi^{out}_{\bk} \ , \nonumber \\
\end{eqnarray}
where $A_{\bk} \ , B_{\bk}$ are Bogolubov
coefficients
\begin{equation}
A_{\bk}=e^{2\pi \omega_{\bk}+\pi i/2}
B^*_{\bk}=
\sqrt{\omega_{\bk}\pi}
e^{\pi \omega_{\bk}-\pi i/4}
\left(\frac{\lambda^{-i\omega_{\bk}}}
{\sinh 2\pi \omega_{\bk}
\Gamma(1-2i\omega_{\bk})}\right) \ . 
\end{equation}
In time-dependent theory we are usually
interested in the number of particles that
are created during the time evolution of given
state. Let us introduce 
following operators 
\begin{eqnarray}
\ma(\bk,t)=\frac{1}{\sqrt{2\Omega_{\bk}(t)}}
\left[\frac{\delta }{\delta \alpha(\bk)}+\Omega_{\bk}
(t)\alpha(\bk)\right] \nonumber \\
\ma^{\dag}(\bk,t)=
\frac{1}{\sqrt{2\Omega_{\bk}(t)}}
\left[-\frac{\delta }{\delta \alpha(-\bk)}+\Omega_{\bk} 
(t)\alpha(-\bk)\right] \nonumber \\
\end{eqnarray}
so that the Hamiltonian can be written as 
\begin{equation}\label{hata}
H(t)=
\int \ik \Omega_{\bk}(t)\left[
\ma^{\dag}(\bk,t)\ma(\bk,t)+\frac{V_{p+1}}{2}\right]
\ , V_{p+1}=(2\pi)^{p+1} \delta_{\bk}(0) \ ,
\end{equation}
where
\begin{equation}\label{npf}
\ma^{\dag}(\bk,t)\ma(\bk,t)=
\frac{1}{2\Omega_{\bk}(t)}\left[
-\frac{\delta^2}{\delta \alpha(-\bk)
\alpha(\bk)}+\Omega^2_{\bk}(t)\alpha(\bk)
\alpha(-\bk)-\Omega_{\bk}(t)(2\pi)^{p+1}
\delta_{\bk}(0)\right] \ .
\end{equation}
Now  (\ref{hata}) can be interpreted as Hamiltonian
for collection of harmonic oscillators with time-dependent
frequencies $\Omega_{\bk}(t)$. Then we see that it is natural to
define an operator of number of  particles with
momentum $\bk$ at time $t$ as
\begin{equation}
N_{\bk}(t)=\ma^{\dag}(\bk,t)\ma(\bk,t) \ .
\end{equation}
We would like to calculate its 
expectation value for the vacuum wave functional $\Psi^{in} [\phi,t]$.
Generally, the vacuum expectation value of any arbitrary
operator is defined by
\begin{equation}
\left<\mathcal{O}\right>=
\bra{\Psi_0(t)}\mathcal{O}\ket{\Psi_0(t)}=
\int d\alpha(\bk)\Psi^*_0\mathcal{O}
\Psi_0 \ .
\end{equation}
Evaluation of vacuum expectation values 
is easily performed by
 introducing
the source term in the vacuum probability density
as follows
\begin{equation}
|\Psi|_0^2[j]\equiv
|\Psi|^2_0\exp\left[\int \ik \alpha(\bk)j(\bk)\right] \ ,
\end{equation}
where
the vacuum probability
density is 
\begin{equation}
|\Psi|^2_0=\Psi^*_0[\alpha,t]\Psi_0[\alpha,t]=
N_0^2\exp\left\{-\frac{1}{2}
\int \ik
2\alpha(-\bk)\tg_R (\bk)
\alpha(\bk)\right\}  \ .
\end{equation}
Then we  get
\begin{eqnarray}\label{vacalpha}
\left<\alpha(\bk)\alpha(\bk')\right>=
\left.\frac{\delta^2}{\delta j(\bk)
\delta j(\bk')}\left<\Psi_0|\Psi_0\right>_{J}
\right|_{J=0}=|N_0(t)|^2\times\nonumber \\
\times\left.\frac{\delta^2}{\delta j(\bk)
\delta j(\bk')}\int
d\alpha\exp\left[-\frac{1}{2}\int
\ikkk\left(2\tg_R(\bk'',t)|\alpha(\bk'')|^2
 -2j(\bk'')\alpha(\bk'')\right)
\right]\right|_{J=0}=
\nonumber \\
=\left.\frac{\delta^2}{\delta j(\bk)
\delta j(\bk')}\exp
\left[\frac{1}{2}\int \ikkk\frac{(2\pi)^{2(p+1)}
j(\bk'')j(-\bk'')}{2\tg_R(\bk'')}\right]
\right|_{J=0}=
\frac{(2\pi)^{p+1}\delta(\bk+\bk')
}{2\tg_R(\bk',t)} \ . 
\nonumber \\
\end{eqnarray}
In order to evaluate 
\begin{equation}\label{delta2}
\left<\frac{\delta^2}{
\delta \alpha(\bk)\delta \alpha(\bk')}
\right>
\end{equation}
it is convenient to use
integration by parts to
write 
\begin{equation}\label{part}
\Psi_0^*\frac{\delta^2}{
\delta \alpha(\bk)\delta \alpha(\bk')}\Psi_0=
-\frac{\delta}{\delta \alpha(\bk')}\Psi^*_0
\frac{\delta}{\delta \alpha(\bk')}\Psi_0=
-\tg^*(\bk',t)\alpha(\bk')\tg(\bk,t)\alpha(\bk) \ ,
\end{equation}
where we have neglected the surface term
and we also use the fact that
\begin{eqnarray}
\frac{\delta}{\delta \alpha(\bk)}
\Psi_0=-\tg(\bk,t)\alpha(\bk)\Psi_0 \ .
\nonumber \\
\end{eqnarray}
From (\ref{part}) and from (\ref{vacalpha}) we
get
\begin{equation}
\left<\frac{\delta^2}{
\delta \alpha(\bk)\delta \alpha(\bk')}
\right>=-\frac{\tg^*(\bk,t)
\tg(\bk,t)}{2\tg_R(\bk,t)}
(2\pi)^{p+1}\delta(\bk+\bk') \  .
\end{equation}
Then the vacuum expectation value of the
number of particles with momentum $\bk$ is equal to
\begin{eqnarray}
\left<N(\bk,t)\right>=\left<
\frac{1}{2\Omega_{\bk}}\left[
-\frac{\delta^2}{\delta \alpha(-\bk)
\alpha(\bk)}+\Omega^2_{\bk}(t)\alpha(\bk)
\alpha(-\bk)-\Omega_{\bk}(t)(2\pi)^{p+1}
\delta_{\bk}(0)\right]\right>=\nonumber \\
=\frac{(2\pi)^{p+1}\delta_{\bk}(0)}{2\Omega_{\bk}(t)}
\left(\frac{\tg^*(\bk,t)
\tg(\bk,t)}{2\tg_R(\bk,t)}
+
\frac{\Omega^2_{\bk}(t)}{2\tg_R(\bk',t)}-\Omega_{\bk}(t)\right)
=\nonumber \\
=(2\pi)^{p+1}\delta_{\bk}(0)
\frac{\left(\Omega_{\bk}(t)-\tg(\bk,t)\right)
\left(\Omega_{\bk}(t)-\tg^*(\bk,t)\right)
}{2\Omega_{\bk}(t)(\tg(\bk,t)+\tg^*(\bk,t))} \ .
 \nonumber \\
\end{eqnarray}
Since the factor $(2\pi)^{p+1}\delta_{\bk}(0)$ is equal
to the volume of spatial section of  D(p+1)-brane it is convenient
to define the vacuum
expectation value of the spatial
density of the number of particle with momentum $\bk$  as
\begin{equation}\label{Nden}
\left<\mathcal{N}_{\bk}(t)\right>\equiv
\frac{\left<N_{\bk}(t)\right>}{V}=
\frac{\left(\Omega_{\bk}(t)-\tg(\bk,t)\right)
\left(\Omega_{\bk}(t)-\tg^*(\bk,t)\right)
}{2\Omega_{\bk}(t)(\tg(\bk,t)+\tg^*(\bk,t))} \ .
\end{equation}
Let us calculate $\left<\mathcal{N}_{\bk}(t)\right>$ 
for  the vacuum state functional
$\Psi^{in}[\phi,t]$ in the limit $t\rightarrow \infty$.
Using
\begin{eqnarray}\label{1}
\frac{\dot{\psi}_{\bk}^{in*}}
{\psi_{\bk}^{in*}}=
\frac{A_{\bk}\dot{\psi}_{\bk}^{out*}
-B^*_{\bk}\dot{\psi}_{\bk}^{out}}
{A_{\bk}\psi_{\bk}^{out*}
-B^*_{\bk}\psi_{\bk}^{out}}
\nonumber \\
\end{eqnarray}
and the asymptotic behaviour of
$\psi^{out}_{\bk}$ for $t\rightarrow
\infty$
\begin{equation}
\dot{\psi}_{\bk}^{out}
\sim
-i\sqrt{\lambda}
e^{t/2}\psi_{\bk}^{out} \ , \mathrm{for}  \  t\rightarrow \infty
\end{equation}
we obtain an asymptotic form of
the kernel $\tilde{G}^{in}(\bk,t)$
\begin{eqnarray}
\tg^{in}(\bk,t)=-i\frac{\dot{\psi}_{\bk}^{in*}}
{\psi_{\bk}^{in*}}=
\sqrt{\lambda}e^{t/2}
\frac{\left(1+\gamma_{\bk}e^{-4i\sqrt{\lambda}e^{t/2}}
\right)}{\left(1-\gamma_{\bk}e^{-4i\sqrt{\lambda}e^{t/2}}
\right)} \  \mathrm{for} \ t\rightarrow \infty \ , \nonumber \\
\end{eqnarray}
where
\begin{equation}
\gamma_{\bk}=\frac{B^*_{\bk}}{A_{\bk}} \ .
\end{equation}
For $t\rightarrow \infty$ we also have $\Omega_{\bk}=\sqrt{\lambda}e^{t/2}$
and consequently the vacuum expectation
value of  the density 
particles with momentum $\bk$   
on half S-brane  at far future is equal to 
\begin{equation}
\left<\mathcal{N}_{\bk}(\infty)\right>=
\frac{|\gamma_{\bk}|^2}
{1-|\gamma_{\bk}|^2}=
|B_{\bk}|^2 \ .
\end{equation}
Now using the fact that
\begin{eqnarray}
B_{\bk}^*=\sqrt{\omega_{\bk}\pi}
e^{-\pi \omega_{\bk}}e^{-\frac{3\pi i}{4}}
\frac{\lambda^{-i\omega_{\bk}}}{
\sinh 2\pi\omega_{\bk}
\Gamma(1-2i\omega_{\bk})} \ , 
\nonumber \\
|\Gamma(1+2i\omega_{\bk})|^2=
\frac{2\pi \omega_{\bk}}{\sinh(2\pi \omega_{\bk})} \ , 
\nonumber \\
|B_{\bk}|^2=\frac{e^{-2\pi\omega_{\bk}}}{
2\sinh 2\pi \omega_{\bk}} 
=\frac{1}{e^{\frac{\omega_{\bk}}{T_H}}-1}  \ ,
T_H=\frac{1}{4 \pi} \ 
\nonumber \\
\end{eqnarray} 
we obtain the final result
\begin{equation}
\left<\mathcal{N}_{\bk}(t)\right>=
\frac{1}{e^{\frac{\omega_{\bk}}{T_H}}-1}  \ ,
T_H=\frac{1}{4 \pi} \ .
\end{equation}
We see  that even if $\Psi^{in}[\phi,t]$ is pure
state  the density of particles with momentum $\bk$ at far future
is the same as the density of particles
 in the thermal state at the temperature $T_H=\frac{1}{4\pi}$.
 According to \cite{Strominger:2002pc}
this means that the branes tries to produce open strings
at the Hagedorn temperature although the final state is
pure state. The fact that this "temperature" is so high
means that $g_s$ corrections are important even for
$g_s\rightarrow 0$.  As was stressed in
previous section, we will follow \cite{Maloney:2003ck}
and we will not consider these in our description.
\section{Full S-brane }\label{fourth}
In this section we formulate Schr\"odinger
picture description of the quantum field
theory on the full S-brane.  In fact, the
 analysis is almost 
the same as in the previous section. The only
difference is that modes $\psi_{\bk}$ in (\ref{kersol}) obey
following differential equation
\begin{equation}
\partial_t^2\psi_{\bk}+2\lambda \cosh t \psi_{\bk}
+\omega_{\bk}^2\psi_{\bk}\equiv
\left(\partial_t^2+\Omega^2_{\bk}(t)\right)\psi_{\bk}=0 \ .
\end{equation}
This equation  was 
studied in the context of S-brane dynamics  in
 \cite{Maloney:2003ck}.
 According to \cite{Maloney:2003ck}
 solution $\psi_{\bk}^{in}(t)$ that approaches
positive frequency mode at far past 
and  $\psi_{\bk}^{out}(t)$ that
approaches positive frequency modes
at the far future  are given
\begin{eqnarray}
\psi_{\bk}^{in}(t)=\sqrt{\frac{\pi}{2}}
(ie^{2\pi \tilde{\omega}_{\bk}})^{1/2}
H^{(1)}(-2i\tilde{\omega}_{\bk},-t/2)
\ , \nonumber \\
\psi_{\bk}^{out}(t)=\sqrt{\frac{\pi}{2}}
(ie^{2\pi \tilde{\omega}_{\bk}})^{-1/2}
H^{(2)}(-2i\tilde{\omega}_{\bk},t/2)
\ . \nonumber \\
\end{eqnarray}
The explicit form of the functions $H^{(1)} \ ,
H^{(2)}$ can be found in  \cite{Maloney:2003ck}.
For our purposes it is sufficient to know
their asymptotic behaviour 
\begin{eqnarray}
H^{(1)}(-2i\tilde{\omega}_{\bk},
t/2) \rightarrow \frac{\lambda^{-1/4}}{\sqrt{\pi}}
e^{-\pi\tilde{\omega}_{\bk}}\exp \left(-\frac{t}{4}+2i\sqrt{\lambda}e^{t/2}
-i\frac{\pi}{4}\right) \ , \mathrm{as} \ t\rightarrow \infty  \ , 
\nonumber \\
H^{(2)}(-2i\tilde{\omega}_{\bk},
t/2) \rightarrow \frac{\lambda^{-1/4}}{\sqrt{\pi}}
e^{\pi\tilde{\omega}_{\bk}}
\exp \left(-\frac{t}{4}+2i\sqrt{\lambda}e^{t/2}
-i\frac{\pi}{4}\right) \ , \mathrm{as} \ t\rightarrow \infty \ .
\nonumber \\
\end{eqnarray}
The frequency $\tilde{\omega}_{\bk}$ is
function of $\omega_{\bk}  \ \mathrm{and} \ \lambda $
which for small $\lambda$ can be written as
\begin{equation}
\omega^2_{\bk}=\tilde{\omega}_{\bk}^2+
\frac{2\lambda^2}{4\tilde{\omega}^2_{\bk}+1}+
\frac{(20\tilde{\omega}^2_{\bk}-7)\lambda^4}
{2(4\tilde{\omega}^2_{\bk}+1)^3(\tilde{\omega}^2_{\bk}+1)}
+\dots \ .
\end{equation}
The relation between in and out modes is
\begin{eqnarray}
\psi^{in}_{\bk}(t)=\frac{1}{2\sinh 2\pi \tilde{\omega}_{\bk}}
\left[i\left(e^{2\pi \tilde{\omega}_{\bk}}
\xi_{\bk} -\frac{e^{-2\pi\tilde{\omega}_{\bk}}}{\xi_{\bk}}
\right)\psi^{out}_{\bk}(t)+
\left(\xi_{\bk}-\frac{1}{\xi_{\bk}}\right)\psi^{out *}_{\bk}(t)
\right]=
\nonumber \\
=\alpha_{\bk}\psi^{out}_{\bk}(t)+
\beta_{\bk} \psi_{\bk}^{out *}(t) \ , \nonumber \\
\end{eqnarray}
where $\alpha_{\bk} \ , \beta_{\bk}$ are 
the Bogolubov coefficients
\begin{equation}
\alpha_{\bk}=\frac{i}
{2\sinh 2\pi \tilde{\omega}_{\bk}}
\left(e^{2\pi \tilde{\omega}_{\bk}}
\xi_{\bk} -\frac{e^{-2\pi\tilde{\omega}_{\bk}}}{\xi_{\bk}}
\right) \ ,
\beta_{\bk}=
\frac{1}
{2\sinh 2\pi \tilde{\omega}_{\bk}}
\left(\xi_{\bk}-\frac{1}{\xi_{\bk}}\right) \ . 
\end{equation}
Although the dependence of $\tilde{\omega}_{\bk}
\ , \xi_{\bk}$ on $\omega_{\bk}\ ,\lambda$ is in general
quite complicated, it can be shown that
Bogolubov coefficients satisfy the unitarity
relation $|\alpha_{\bk}|^2-|\beta_{\bk}|^2=1$.
In the case of large $\omega_{\bk}$ and
$\lambda\ll \omega_{\bk}$ we have
\begin{equation}
\tilde{\omega}_{\bk}=\omega_{\bk}[
1+O(\lambda^2/\omega^4_{\bk})] \ ,
\xi_{\bk}=\frac{\Gamma(1-2i\omega_{\bk})}
{\Gamma(1+2i\omega_{\bk})}\lambda^{2i\omega_{\bk}}
[1+O(\lambda^2/\omega_{\bk}^2)]
\end{equation}
and the Bogolubov coefficients are
\begin{equation}
\alpha_{\bk}=\frac{\sin(\theta_{\bk}+2\pi i\omega_{\bk})}
{\sinh 2\omega_{\bk}} \ ,
\beta_{\bk}=-i\frac{\sin \theta_{\bk}}
{\sinh 2\pi \omega_{\bk}} \ ,
e^{i\theta_{\bk}}=
\lambda^{-2i\omega_{\bk}}
\frac{\Gamma(1+2i\omega_{\bk})}
{\Gamma(1-2i\omega_{\bk})} \ .
\end{equation}
We will be mainly interested in the case when
 $\lambda \rightarrow 0$ which has nice physical
interpretation. For such a $\lambda$ 
there is a long region around $t=0$ of
duration $\ln \lambda$ in which the 
interaction can be neglected and we just have
ordinary unstable brane. There
is then a natural    vacuum state $\Psi^E[\phi,t]$
which is defined by the requirement that
there   are no particles at $t=0$. It
is associated with the  function $\psi_{\bk}$ appearing in
the kernel $\tg(\bk,t)=-i\frac{\dot{\psi}_{\bk}}{\psi_{\bk}}$
\begin{equation}
\psi^0_{\bk}(t)=\sqrt{\frac{4\pi \xi_{\bk}}{\sinh 2\pi
\tilde{\omega_{\bk}}}}J(-2i\tilde{\omega}_{\bk},
t/2) .
\end{equation}
Using the relation 
\begin{equation}
J=\frac{1}{2}[H^{(1)}+H^{(2)}]
\end{equation}
$\psi^0_{\bk}$ can be expressed in terms 
of $\psi^{in}_{\bk}$ as
\begin{eqnarray}
\psi^0_{\bk}(t)=a_{\bk}\psi_{\bk}^{in}(t)
-b^*_{\bk}\psi^{in*}_{\bk}(t) \ , 
\nonumber \\
\end{eqnarray}
where
\begin{equation}
a_{\bk}=\frac{e^{\pi \tilde{\omega}_{\bk}-i\frac{\pi}{4}}}
{\sqrt{2\xi_{\bk}
\sinh 2\pi \tilde{\omega}_{\bk}}}
 \ ,
b_{\bk}=-e^{-\pi 
\tilde{\omega}_{\bk}-i\frac{\pi}{4}}
\sqrt{\frac{\xi_{\bk}}{2
\sinh 2\pi \tilde{\omega}_{\bk}}} \ .
\end{equation}
We can easily confirm that the vacuum wave functional
$\Psi^E[\phi,t]$ does not contain particles at  $t=0$
which implies that it is correct vacuum state for unstable
D-brane at $t=0$. In fact, for $\lambda\rightarrow 0$
we  have
\begin{equation}
\Omega_{\bk}(0)=
\sqrt{\omega_{\bk}^2+\lambda^2}\sim
\omega_{\bk}(1+\lambda^2/\omega^2_{\bk})
\ 
\end{equation} 
while the kernel $\tilde{G}^E(\bk,t)$ 
around $t=0$ is equal to 
\begin{equation}
\tilde{G}(\bk,t)=
-i\frac{\dot{\psi}^{0*}_{\bk}(t)}
{\psi^{0*}_{\bk}(t)}\sim \tilde{\omega}_{\bk}
\sim \omega_{\bk} \ 
\end{equation}
and hence the vacuum state functional $\Psi^E[\phi,t]$
at $t=0$ has the form
\begin{equation}\label{unstD}
\Psi^E[\phi,0]=N\exp\left[
-\frac{1}{2}\int \ik \alpha(-\bk)
\omega_{\bk}\alpha(\bk)\right] \ 
\end{equation}
which is standard Minkowski  vacuum wave functional. 
Following  calculations presented in the
previous section it is easy to see that particle excitations 
are absent in the state (\ref{unstD}).

We can also calculate the number of particles created
during  the time evolution of the Euclidean vacuum 
state $\Psi^E[\phi,t]$. In fact, the calculation  is the same
as in case of half S-brane studied in the previous
section so that the vacuum expectation value of  density of particles 
with momentum $\bk$ in the state $\Psi^E$ 
at the asymptotic future is equal to
\begin{equation}
\left<\mathcal{N}_{\bk}(\infty)\right>=
\frac{|\gamma_{0\rightarrow  out , \bk}|^2}
{1-|\gamma_{0\rightarrow  out ,  \bk}|^2}=
|b_{\bk}|^2
 \ . 
\end{equation}
As in case of half S-brane we see that the vacuum
state $\Psi^E[\phi,t]$ despite the fact that it is
pure state it is populated  in far future by particles with the
particle density corresponding to the thermal state
with the temperature $T_H=\frac{1}{4\pi}$. 
In the same way we can calculate 
the vacuum expectation value of particle density
in the asymptotic past and we again obtain the
result that the particle occupation number is
thermal with the temperature $T_H=1/4\pi$.  

So far we have studied the pure
states of quantum field theory on S-brane. In the
next section we will consider more general approach
when we will construct mixed states in
the Schr\"odinger picture description.
\section{S-brane thermodynamics in
Schr\"odinger picture}\label{fifth}
In this section we will formulate
the Schr\"odinger picture description of  
S-brane thermodynamics.
When the initial state of a system is a pure state,
described by definite wave functional $\Psi$, the 
time-dependent Schr\"odinger equation determines
uniquely  its time evolution. This  situation
 was studied in the previous two  sections.
However we can consider more general case when
initially at time $t_0$ the system is in mixed state, described
by a functional density matrix $\hat{\rho}(t_0)$.
In the Schr\"odinger picture, when operators
$\hf(\bx) \ , \hp(\bx)$ are time independent while  states change with time,
the Liouville equation that determines the time
evolution of density matrix $\hat{\rho}(t)$, reads
\begin{equation}\label{liueq}
i\frac{\partial \hat{\rho}(t)}{\partial t}=
[\hat{H}(t),\hat{\rho}(t)] \ .
\end{equation}
A formal solution of this equation may be written
in terms of the time evolution operator
\begin{equation}\label{foru}
\hat{U}(t,t')=\mathrm{P}\exp\left[-i\int_{t'}^tdt''
\hat{H}(t'')\right] \  
\end{equation}
in the form
\begin{equation}
\hat{\rho}(t)=\hat{U}(t,t_0)
\hat{\rho}(t_0)\hat{U}^{-1}(t,t_0) \ . 
\end{equation}
In (\ref{foru}) the symbol $\mathrm{P}$ means
path ordering.  
The quantity $\hat{\rho}(t_0)$ determines 
the initial state of the system. Usual choice
is to let the initial state be one
of local thermodynamic equilibrium at a
temperature $T=1/\beta$. The unnormalised
initial density matrix then takes the form
\begin{equation}
\hat{\rho}(t_0)=\exp \left(-\beta \hat{H}(t_0) 
\right) \ .
\end{equation}
Now the ensemble averages of operators are given
by the expression
\begin{equation}
\left<\hat{O}(t)\right>=
\frac{\tr\left[\hat{\rho}(t)\hat{O}\right]}
{\tr \hat{\rho}(t)}=\frac{
\tr\left[ \hat{\rho}(t_0)\hat{U}(t_0,t')
\hat{U}(t',t)\hat{O}\hat{U}(t,t_0)\right]}
{\tr \hat{\rho}(t_0)} \ ,
\end{equation}
where we have used cyclicity of the trace and
 we have also inserted the identity
$\hat{U}(t,t')\hat{U}(t',t)$. 
Since the spatial sections are flat 
we perform Fourier transform of 
$\phi(\bx)=\int \ik e^{i\bk\bx}$ and 
 consider the density
matrix in the form
\begin{eqnarray}\label{rhoalpha}
\rho(\phi_1,\phi_2,t)=
\prod_{\bk}
\rho_{\bk}(\alpha_1(\bk),\alpha_2(\bk),t) \ ,
\nonumber \\
\rho(\alpha_1(\bk),\alpha_2(\bk),t)=N_{\bk}(t)
\exp\left\{-\frac{1}{2}A_{\bk}(t)\alpha_1(\bk)
\alpha_1(-\bk)-\frac{1}{2}A^*_{\bk}(t)
\alpha_2(\bk)\alpha_2(-\bk)-\right. \nonumber \\
\left.-B_{\bk}(t)\alpha_1(\bk)\alpha_2(-\bk)
\right\} \ , \nonumber \\
\end{eqnarray}
where $B_{\bk}(t)$ is real as follows from
the  Hermicity of the density matrix 
$\rho(\phi_1,\phi_2)=
\rho^*(\phi_2,\phi_1)$. 
Now the equations that
determine time evolution of  coefficients $A_{\bk}(t) \ ,
B_{\bk}(t)$, follow from (\ref{liueq}) that
in k-space 
 has the form
\begin{eqnarray}
i\frac{\partial \rho(\alpha_1,
\alpha_2,t)}{\partial t}=
\int \ik\left[-\frac{1}{2}\left(\frac{\delta^2}{
\delta \alpha_1(\bk)\alpha_1(-\bk)}
-\frac{\delta^2}{
\delta \alpha_2(\bk)\alpha_2(-\bk)}\right)+\right.
\nonumber \\
\left.+\frac{\Omega^2_{\bk}(t)}{2}\left(\alpha_1(\bk)
\alpha_1(-\bk)-
\alpha_2(\bk)\alpha_2(-\bk)\right)\right]
\rho(\alpha_1,\alpha_2,t)  \ . \nonumber \\
\end{eqnarray}
Since the modes do not mix the equations
for kernels in the density matrix are obtained
by comparing the powers of $\alpha$ on
both sides of the above equation which
gives  following equations
for coefficients $A_{\bk}\ , B_{\bk}$
\begin{eqnarray}\label{coeq}
\frac{\dot{N}_{\bk}}{N_{\bk}}=
-i\left(A_{\bk}-A^*_{\bk}\right) \ , \nonumber \\
\dot{A}_{\bk}=-i\left[A_{\bk}^2-
B_{\bk}^2+\Omega^2_{\bk}\right] \nonumber \\
\dot{B}_{\bk}=-iB_{\bk}(A_{\bk}-
A^*_{\bk}) \ . \nonumber \\
\end{eqnarray}
One  can show that
\begin{eqnarray}\label{twrel}
\frac{d}{dt}\left(\frac{B_{\bk}(t)}
{A_{\bk R}(t)}\right)=0 \nonumber \\
\frac{d}{dt}\left(\frac{N_{\bk}(t)}
{\sqrt{A_{\bk R}(t)+B_{\bk}(t)}}\right)=0  \ , \nonumber \\
\end{eqnarray}
where $A_{\bk R}=\frac{1}{2}(A_{\bk}+
A^*_{\bk}) \ , 
A_{\bk I}=\frac{1}{2i}(A_{\bk}-A_{\bk}^*)$.
The relations (\ref{twrel}) suggest that we can 
take
\begin{equation}
\mathcal{A}_{\bk R}(t)=
C_{\bk}A_{\bk R}(t)=D_{\bk}B_{\bk}(t) \ ,
\mathcal{A}_{\bk I}(t)=A_{\bk I}(t) \ , 
\end{equation}
where the constants $C_{\bk}, D_{\bk}$
depend on choice of initial conditions. 
We can also define following quantity  
$\mathcal{A}_{\bk}=
A_{\bk R}+i\mathcal{A}_{\bk I}$. From
(\ref{coeq}) immediately follows that
 $ \mathcal{A}_{\bk}$ 
obeys equation
\begin{equation}
\dot{\mathcal{A}}_{\bk}(t)=-
i\left[\mathcal{A}^2_{\bk}(t)-\Omega^2_{\bk}(t)
\right] \ .
\end{equation}
This equation can be linearised as
\begin{equation}
\mathcal{A}_{\bk}(t)=-i\frac{\dot{f}_{\bk}(t)}{f_{\bk}(t)}  \ , 
\end{equation}
where $f_{\bk}$ obeys
\begin{equation}
\ddot{f}_{\bk}(t)+\Omega^2_{\bk}(t)f_{\bk}(t)=0 \ .
\end{equation}
In the previous equation  $\Omega^2_{\bk}(t)=\omega_{\bk}^2+
\lambda e^t$ for half S-brane and
$\Omega^2_{\bk}(t)=\omega_{\bk}^2+2\lambda
\cosh t$ for full S-brane and hence the modes
$f_{\bk}(t)$ are the same as $\psi_{\bk}(t)$ given
in previous sections. From the  known functions 
 $f_{\bk}(t)$ we immediately get
\begin{eqnarray}
A_{\bk R}(t)=\frac{1}{2}
(A_{\bk}(t)+A_{\bk}^*(t))=
\frac{1}{C_{\bk}|f_{\bk}|^2} \ ,
\nonumber \\
A_{\bk I}=
\frac{1}{2i}(\mathcal{A}_{\bk}-
\mathcal{A}^*_{\bk})
=-\frac{1}{2}
\frac{\dot{f}_{\bk}f_{\bk}^*+
\dot{f}_{\bk}^*f_{\bk}}{|f_{\bk}|^2} \ ,  \nonumber \\
B_{\bk}(t)=\frac{1}{D_{\bk}|f_{\bk}|^2}  \ .
\nonumber \\
\end{eqnarray}
Now we are ready to calculate the equal-time
Green functions 
\begin{eqnarray}
G(\bx,\by,t)=\tr \hat{\rho}(t)\hf(\bx)\hf(\by)=
\int d\phi \rho(\phi,\phi)\phi(\bx)\phi(\by)=
\nonumber \\ 
=\int \ik\ikk 
e^{i(\bk\bx+\bk'\by)}
\left<\alpha(\bk)\alpha(\bk')\right> \ , \nonumber \\
\end{eqnarray}
where 
\begin{equation}\label{twof}
\left<\alpha(\bk)\alpha(\bk')\right>=
\int d\alpha \rho(\alpha,\alpha)
 \alpha(\bk)\alpha(\bk') \ .
\end{equation}
As in section (\ref{third}) we define
\begin{equation}
\rho(J, \,\alpha,\alpha)\equiv
\rho(\alpha,\alpha)\exp \left(\int
\ik J(\bk)\alpha(\bk)
\right) \ . 
\end{equation}
Since the diagonal form of the
 density matrix is equal to
\begin{eqnarray}
\rho(\alpha,\alpha,t)=
\prod_{\bk}\rho_{\bk}(\alpha,\alpha,t)=
\exp\left[-\int \ik
\left(A_{\bk R}+B_{\bk R}\right)\alpha(\bk)
\alpha(-\bk)\right] \nonumber \\
\end{eqnarray}
we  can  easily  calculate two-point function 
(\ref{twof})
\begin{eqnarray}
\left<\alpha(\bk)\alpha(\bk')\right>=
\frac{\delta^2}{\delta J(\bk)
\delta J(\bk')}\int d\alpha N(t)
\exp\left\{-\frac{1}{2}\int \ik\left[
2\left(A_{\bk  R}+B_{\bk R}\right)|\alpha(\bk)|^2
+\right.\right. \nonumber \\
\left.\left.+2J(\bk)\alpha(\bk)\right]\right\}
=\frac{\delta^2}{\delta J(\bk)
\delta J(\bk')}\exp\left[
\frac{1}{2}\int d\bk''\frac{(2\pi)^{p+1}J(\bk'')
J(-\bk'')}{2(A_{\bk R}+B_{\bk R})}\right]
=\nonumber \\
=(2\pi)^{p+1}\frac{\delta(\bk+\bk')}{2
(A_{\bk R}(t)+B_{\bk,R}(t))}=
(2\pi)^{p+1}\frac{\delta(\bk+\bk')|f_{\bk}(t)|^2}{2}
\frac{C_{\bk}D_{\bk}}{C_{\bk}+D_{\bk}} \ . 
 \nonumber \\
\end{eqnarray}
We  choose the constants $C_{\bk} \ ,
D_{\bk}$ in such a way  so that the density matrix
$\hat{\rho}(t_0)$ 
at time $t_0$ 
reduces to the thermal density matrix 
at the temperature $T=1/\beta$.
This  can be achieved with these constants
\begin{eqnarray}\label{inv}
C_{\bk}=2\tanh(\beta \omega_{\bk}) \ ,
\nonumber \\
D_{\bk}=-2\sinh(\beta \omega_{\bk}) \ ,
\nonumber \\
f_{\bk}(t_0)=\frac{1}{\sqrt{2\omega_{\bk}}} \ ,
\dot{f}_{\bk}(t_0)=
-i\omega_{\bk}f_{\bk} \ , \nonumber \\
N_{\bk}(t_0)=\left[\frac{\omega_{\bk}}{\pi}
\tanh \left(\frac{\beta \omega_{\bk}}
{2}\right)\right] \ \nonumber \\
\end{eqnarray}
and consequently  the equal time two point function is
\begin{equation}
\left<\alpha(\bk)\alpha(\bk')\right>
=(2\pi)^{p+1}\delta(\bk+\bk')
\frac{|f_{\bk}(t)|^2}{2}\coth\left(
\frac{\beta \omega_{\bk}}{2}\right) \ .
\end{equation}
In the same way we obtain
\begin{equation}\label{rhop}
\left<\frac{\delta^2}{\delta \alpha_1
(\bk)\delta \alpha_2(\bk')}\right>=
(2\pi)^{p+1}\delta(\bk+\bk')\frac{A_{\bk}(t)A^*_{\bk}(t)-
B_{\bk}^2(t)}{2(A_{\bk R}(t) +B_{\bk}(t))} \ . 
\end{equation}
As it is clear from previous sections it
is natural to take $t_0$ to be equal to
$-\infty$ for half S-brane, while 
for full S-brane  $t_0=0$. 
At time $t=t_0$ we  have
\begin{eqnarray}
\left<\alpha 
(\bk)\alpha(\bk')\right>=
(2\pi)^{p+1}\delta(\bk+\bk')
\frac{1}{2\omega_{\bk}}\coth\left(\frac{\beta\omega_{\bk}}
{2}\right) \ , \nonumber \\
\left<\frac{\delta^2}{\delta \alpha 
(\bk)\delta \alpha(\bk')}\right>=
-\omega_{\bk}^2(2\pi)^{p+1}\delta(\bk+\bk')
\frac{\omega_{\bk}}{2}\coth\left(\frac{\beta\omega_{\bk}}
{2}\right) \ . \nonumber \\
\end{eqnarray}
It is easy to see that  the density of particles with momentum 
$\bk$ at time $t_0$ is equal to
\begin{eqnarray}
\mathcal{N}_{\bk}(t_0)=\frac{1}{V}\tr
\hat{N}_{\bk}\hat{\rho}(t_0)
=\frac{1}{
\exp\left(\beta\omega_{\bk}\right)-1} \ . 
\nonumber \\
\end{eqnarray}
as it should be for a system at thermal state with
temperature $T=1/\beta$. 

Above we have constructed mixed states for S-branes that
approach standard thermal states at far past in case
of half S-brane, or at $t=0$ in case of full S-brane. 
These mixed states can be defined for any initial
temperature $T=1/\beta$. However as was shown in
\cite{Maloney:2003ck} for temperatures $T=\frac{1}{2\pi n}$
these mixed states have interesting property that
 Green functions evaluated in these states
 retain their thermal periodicity
for any time $t$. To see this, we  firstly 
show that the evolution 
operator (\ref{foru}) on half S-brane obeys 
\begin{eqnarray}\label{uid}
\hat{U}(t+2\pi i n,t)=
\mathrm{P}\exp\left[-i\int_t^{t+2\pi i n}dt' \hat{H}(t')\right]=
\nonumber \\
=\mathrm{P}
\exp\left[-i\int_t^{t+2\pi i n}dt' \left(\hat{H}(t_0)+
(m^2(t')-m^2)\right)\right]=
\nonumber \\
=\exp\left[2\pi  n\hat{H}(t_0)\right] \ ,
\nonumber \\
\end{eqnarray}
where  we have used
\begin{equation}
\int_t^{t+2\pi i n}dt' (m^2(t')-m^2)=
i\lambda e^t\int_0^{2\pi  n}dy 
e^{iy}=\lambda e^t\left[e^{2\pi i n}-1\right]=0 \ .
\end{equation}
 It is clear that for full
S-brane the integral $\int_t^{t+2\pi i n}dt' (m^2(t)-m_0^2)$
is zero as well. And finally, $\hat{H}(t_0)$ in (\ref{uid}) 
is the time-independent Hamiltonian of the free scalar field
\begin{equation}\label{hamfree}
\hat{H}(t_0)=\int d\bx \left[\hp^2+\delta^{ab}\partial_a
\hf\partial_b\hf+m^2\hf^2\right] \ .
\end{equation}
Using (\ref{hamfree}) we immediately get
\begin{equation}
\hat{U}(t_0+2\pi i n,t_0)=
\exp\left[2\pi  n
\hat{H}(t_0)\right]=\hat{\rho}_n(t_0)^{-1}
\end{equation}
and also
\begin{equation}
U(t_0-2\pi i n,t_0)=
\exp\left[-2\pi  n
\hat{H}(t_0)\right]=\hat{\rho}_n(t_0) \ ,
\end{equation}
where $\hat{\rho}_n(t_0)$ is initial density matrix
at temperature $T=\frac{1}{2\pi n}$. 
For our next purposes it will be useful to
rewrite the equal-time two point function in terms of Heisenberg
picture   operators
\begin{eqnarray}\label{twoH}
G(\bx,\by,t)=\tr \hat{\rho}(t)\hf(\bx)
\hf(\by)=\tr \hat{U}(t,t_0)
\hat{\rho}(t_0)\hat{U}^{-1}(t,t_0)
\hf(\bx)\hat{U}(t,t_0)\times \nonumber \\
\times\hat{U}^{-1}(t,t_0)\hf(\by)\hat{U}(t,t_0)=
\tr \hat{\rho}_H\hf_H(\bx,t)\hf_H(\by,t) \ ,
\nonumber \\
\end{eqnarray}
where
\begin{eqnarray}
\hf_H(\bx,t)=\hat{U}^{-1}(t,t_0)
\hf(\bx)\hat{U}(t,t_0)=
\hat{U}_H(t,t_0)\hf(\bx)
\hat{U}_H^{-1}(t,t_0) \ , \nonumber \\
\hat{\rho}_H=\hat{\rho}(t_0) \ , 
\hat{U}_H(t,t_0)\equiv \mathrm{P}\exp
\left[i\int_{t_0}^tdt' \hat{H}(t')\right] \ . 
\nonumber \\
\end{eqnarray}
Now we generalise the equal-time two point function
written in Heisenberg representation   (\ref{twoH}) 
 to the two point functions evaluated at 
different times  $t,t'$
\begin{equation}\label{twoHd}
G(\bx,t,\bx',t')=
\tr \hat{\rho}_H \hat{\phi}_H(\bx',t')
\hat{\phi}_H(\bx,t) \ . 
\end{equation}
Using the fact that
\begin{eqnarray}
\hat{U}_H(t+2\pi i n,t_0)=
\hat{U}_H(t+2\pi i n,t)
\hat{U}_H(t,t_0)=\nonumber \\
=
\exp\left[ 2\pi n \hat{H}(t_0)\right]
\hat{U}_H(t,t_0)=
\hat{\rho}_{Hn}\hat{U}_H(t,t_0) \nonumber \\
\end{eqnarray}
we can show that  
(\ref{twoH}) 
 retain its thermal
periodicity at all times
\begin{eqnarray}
G_n(\bx,t+2\pi i n,\bx',t')=
\tr \hat{\rho}_{Hn}
\hat{\phi}_H(\bx',t')
\hat{\rho}_{Hn}\hat{\phi}(\bx,t)
\hat{\rho}_{Hn}^{-1}=\nonumber \\
=
\tr \hat{\rho}_{Hn}\hat{\phi}(\bx,t)
\hat{\phi}_H(\bx',t')=
G(\bx',t',\bx,t)  \  \nonumber \\
\end{eqnarray}
and also
\begin{eqnarray}
G(\bx,t,\bx',t'-2\pi i n)=
\tr \hat{\rho}_{Hn}
\hat{\rho}_{Hn}^{-1}\hat{\phi}_H(\bx',t')
\hat{\rho}_{Hn}\hat{\phi}_H(\bx,t)
=\nonumber \\
=
\tr \hat{\rho}_{Hn}\hat{\phi}_H(\bx,t)
\hat{\phi}_H(\bx',t')=
G(\bx',t',\bx,t)  \  \nonumber \\
\end{eqnarray}
so that we  see that $G_n(\bx,t,\bx',t')$
retain their thermal periodicity for finite
$t,t'$. 
In other words, we can say that among the mixed states on
half S-brane that
can be defined for any temperature there
are mixed states with the temperature $T_n=\frac{1}{2\pi n}$
\footnote{More precisely, in the 
exact open string theory description  of S-brane the condition of locality  
in boundary conformal field theory forces
us to consider the temperatures $T_n$ only
\cite{Maloney:2003ck,Suqawara:2003tp}.}
that reduce to usual thermal vacuum at
past infinity while retaining thermal periodicity
at all times. The same result is valid for the
 full S-brane as well. Such thermal states approximate
the quantum states of open strings on an S-brane created
from incoming closed string excitations. 
\section{Conclussion}\label{seventh}
In this paper we have formulated 
the quantum field theory that describes S-branes
in minisuperspace approach in the language
of Schr\"odinger functional formalism. 
As was shown in \cite{Maloney:2003ck}  
the quantum field theory on unstable D-brane
in the presence of rolling tachyon is similar
to the quantum field theory in time-dependent background.
In particular, the specification of vacuum state
on unstable D-brane in rolling tachyon background
is unambiguous and leads to the open strings
production during the D-brane decay 
\cite{Maloney:2003ck}. 

In this paper we studied the quantum field
theory, that arises in
minisuperspace approach to S-brane dynamics,
in the Schr\"odinger formalism.  The reason for
this choice is to avoid the description of the
vacuum as the "no-particle" state in a Fock space
of states generated by the creation operators
of particles defined with respect to a particular
mode decomposition of quantum field. In particular
we tried to avoid "particle interpretation" of
the out vacua in case of half S-brane where according
to standard interpretation of the D-brane
decay there should not exist any open string
perturbative modes. 

We have also studied S-brane thermodynamics in
Schr\"odinger picture. We have explicitly construct
the density matrix for mixed states on half S-brane that
approach standard thermal mixed states at past infinity. We
have shown, according to  \cite{Maloney:2003ck} that
for temperatures $T=\frac{1}{2\pi n}$ these mixed
states retain their thermal periodicity for all times. We have
seen that these mixed states are natural states
for quantum field theory on S-brane and presumably
define low energy effective theory on S-brane.

To summarise, we mean that the Schr\"odinger formalism
for minisuperspace description of 
 S-brane dynamics gives very natural and intuitive picture
of the vacuum state of open string modes during unstable
D-brane decay.  We also hope that the Schr\"odinger
picture description could be useful in the minisuperspace
approach to the rolling tachyon in closed string theory
\cite{Martinec:2003ka,Strominger:2003ct,Schomerus:2003ct}.
\\
\\
{\bf Acknowledgement}
I would like to thank Prof. Ulf Lindstr\"om and
Prof. Ulf Danielsson for their support in my work. 
This work is partly supported by EU contract 
HPRN-CT-2000-00122.
\\

\end{document}